%% file: paper.tex
\documentclass[conference]{IEEEtran}
\IEEEoverridecommandlockouts
\usepackage{cite}
\usepackage{amsmath,amssymb,amsfonts}
\usepackage{algorithmic}
\usepackage{csquotes}
\usepackage[inline]{enumitem}
\usepackage{graphicx}
\usepackage{hyperref}
\usepackage{cleveref}
\usepackage{listings}
\usepackage{siunitx}
\usepackage{textcomp}
\usepackage{todonotes}
\usepackage[normalem]{ulem}
\usepackage{xcolor}
\usepackage{xspace}
\def\BibTeX{{\rm B\kern-.05em{\sc i\kern-.025em b}\kern-.08em
    T\kern-.1667em\lower.7ex\hbox{E}\kern-.125emX}}

\input{macros}

\lstdefinestyle{mystyle}{
    commentstyle=\color{codegreen},
    keywordstyle=\color{magenta},
    numberstyle=\tiny\color{codegray},
    stringstyle=\color{codepurple},
    basicstyle=\ttfamily\footnotesize,
    breakatwhitespace=false,         
    breaklines=true,                 
    captionpos=b,                    
    keepspaces=true,                 
    numbers=left,                    
    numbersep=5pt,                  
    showspaces=false,                
    showstringspaces=false,
    showtabs=false,                  
    tabsize=2,
    otherkeywords={self},
    escapechar=|
}

\crefname{lstlisting}{listing}{listings}
\Crefname{lstlisting}{Listing}{Listings}

\begin{document}

\title{
Interactive Visualization of Protein RINs \\ using NetworKit in the Cloud
\thanks{The work by AvdG was supported by German Research Foundation (DFG) grant GR 5745/1-1. The work by FBT and HM was supported by DFG Collaborative Research Center (CRC) 1404 FONDA.}
}

\author{\IEEEauthorblockN{Eugenio Angriman\IEEEauthorrefmark{1}\IEEEauthorrefmark{2}, Fabian Brandt-Tumescheit\IEEEauthorrefmark{1}\IEEEauthorrefmark{3}, Leon Franke\IEEEauthorrefmark{4} Alexander van der Grinten\IEEEauthorrefmark{1}\IEEEauthorrefmark{5}\\ and Henning Meyerhenke\IEEEauthorrefmark{1}\IEEEauthorrefmark{6}}
\vspace{0.1cm}
\IEEEauthorblockA{\IEEEauthorrefmark{1}Department of Computer Science, Humboldt-Universität zu Berlin, Germany \\
Email: \IEEEauthorrefmark{2}angrimae@hu-berlin.de, \IEEEauthorrefmark{3}brandtfa@hu-berlin.de, \IEEEauthorrefmark{5}avdgrinten@hu-berlin.de, \IEEEauthorrefmark{6}meyerhenke@hu-berlin.de}
\vspace{0.1cm}
\IEEEauthorblockA{\IEEEauthorrefmark{4}Department of Chemistry, Konstanz Research School Chemical Biology, University of Konstanz, Germany\\
Email: leon.franke@uni-konstanz.de}
}

\maketitle

\begin{abstract}
Network analysis has been applied in diverse application domains.
In this paper, we consider an example from protein dynamics,
specifically residue interaction networks (RINs).
In this context, we use
\nwk{} -- an established package for network analysis
-- to build a cloud-based environment
that enables domain scientists to run their visualization
and analysis workflows on large compute servers, without requiring
extensive programming and/or system administration knowledge.
To demonstrate the versatility of this approach, we use it
to build a custom Jupyter-based widget for RIN visualization.
In contrast to existing RIN visualization approaches,
our widget can easily be customized through simple modifications
of Python code, while both supporting a good feature set and providing 
near real-time speed. It is also easily integrated into analysis pipelines
(e.g., that use Python to feed RIN data into downstream machine learning tasks).
\end{abstract}

\begin{IEEEkeywords}
network analysis, cloud computing, graph visualization,
residue interaction networks, molecular dynamics simulations 
\end{IEEEkeywords}


\section{Introduction}

Graphs are essential to model various real-world phenomena such
as social networks, infrastructure networks, web graphs,
or the spread of epidemics,
among other applications~\cite{newman2018networks}.
Network analysis (and graph mining) is an important tool
to extract data from these graphs~\cite{DBLP:books/cu/LeskovecRU14}. To this end, various
toolkits for network analysis have been developed (see \Cref{sec:related-work} for an overview).
In this work, we focus on \nwk, one of the established toolkits in this area,
with a focus on usability (via a Python-based frontend)
and multi-threaded performance (powered by a C++ backend)~\cite{DBLP:journals/netsci/StaudtSM16}.
The Python/C++ architecture makes \nwk well-suited
for our use case: while domain-scientists can use the Python frontend
to build analysis pipelines without extensive programming knowledge,
the C++ core lets \nwk surpass the performance bottlenecks
of pure Python implementations
(however, we do note that \nwk is not unique
in this regard, see \Cref{sec:related-work}).

\nwk's Python-based frontend allows rapid development of
solutions for domain-specific problems, while only requiring a
limited amount of programming expertise.
In this paper, we further lower this barrier of entry by
presenting a cloud-based \nwk installation that does not
require domain scientists to install any packages on their local machines
anymore.
This approach has two substantial advantages:
first, it avoids the need of domain scientists to deal with
the installation of complicated scientific computing packages,
especially when multiple such packages (and not only \nwk) are required.
This task can instead be performed by system administration experts.
Secondly, a cloud-based approach enables domain scientists to run their analysis pipelines on cloud-hosted servers that provide appropriate
resources for computationally intensive tasks
(e.g., large numbers of CPU cores and/or large amounts of RAM).
A cloud-based approach also fosters research to be 
collaborative and shareable.

Based on this cloud-based environment, we consider a use case from
the analysis of molecular dynamics (MD) simulations,
namely the investigation of protein dynamics via residue interaction networks
(RINs, see \Cref{sec:rin} for a definition).
To this end, we extend
\nwk's visualization capabilities
to take domain-specific requirements into account;
as a result, we present a Jupyter notebook widget that can be used
to visualize RINs interactively (see \Cref{sec:new-viz}).
Since expensive recomputations of the RIN layout are delegated to a
cloud server, our widget updates in real-time, even when
computationally expensive analytics (\eg centrality measures such as
betweenness and closeness, or community detection) are incorporated into the
visualization. 

\paragraph*{Outline and contribution}
We first give an overview of \nwk, including recent changes to its ecosystem (in \Cref{sec:nwk-overview}).
Next, we present the setup of our \nwk-based cloud deployment (in \Cref{sec:cloud}) and how we use it to 
provide a low barrier of entry for non-experts in
network analysis.
To the best our knowledge, there are no other
Python-programmable cloud environments available yet
that explicitly support network analysis in domain-specific applications.
After an overview over a relevant domain-specific problem concerning the analysis of RINs (in \Cref{sec:rin}), we show how \nwk
helps to solve application-specific questions in this domain
by providing customized visualization tools (in \Cref{sec:new-viz}).
In contrast to existing RIN visualization solutions,
our use of \nwk on a cloud server enables domain scientists to interactively explore entire simulation data sets and their graph-based features in real time in a Python-based cloud environment.


\section{\nwk Overview}
\label{sec:nwk-overview}

The network analysis package \nwk was first released in 2013.
Its architecture was previously described in Refs.~\cite{DBLP:journals/netsci/StaudtSM16}
and~\cite{networkit_spp};
here we only give a brief overview.
\nwk offers numerous unique algorithms that focus
on handling large input graphs.
These algorithms cover different families of popular network analysis problems such as
distance computations (in the \texttt{distance} module), community detection (\texttt{community}),
network centrality (\texttt{centrality}), network generators (\texttt{generators}),
connected components (\texttt{components}), and others.

For many network analytic problems, approximation is often the only feasible technique
to obtain results in reasonable time.
Hence, many of \nwk's modules implement approximation algorithms.
For example, most centrality measures
can be computed either exactly for small to medium networks or approximated for
larger networks.

We use Cython as a glue between C++ and Python code (\href{https://cython.org/}{cython.org}).
Shared-memory parallelism (via OpenMP) is used extensively throughout the C++ code base.

\subsection{Improvements on Previous \nwk Work}

We briefly report selected features that were added to \nwk since
previous reports~\cite{networkit_spp,DBLP:journals/netsci/StaudtSM16}.\footnote{
We remark that some of the features mentioned in this section were developed in a
collaborative effort with contributors external to the \nwk project.}
Additions to the \texttt{centrality} module
include a normalization strategy for PageRank based on Ref.~\cite{DBLP:conf/www/BerberichBWV07}
that makes PageRank scores comparable across different graphs.
New community detection algorithms include: a parallel version of the Leiden algorithm
~\cite{DBLP:journals/corr/abs-1810-08473}, a parallel version
of the Louvain algorithm based on map equation~\cite{bohlin2014community},
and the algorithm by McDaid \etal~\cite{DBLP:journals/corr/abs-1110-2515} to compute the Normalized
Mutual Information, a widely used similarity measure to compare pairs of communities.
These community detection algorithms can easily be integrated into
the RIN visualization widget that we develop in \Cref{sec:new-viz}.
For a description of a new graph drawing module based on
\texttt{Plotly} (\url{https://github.com/plotly}), see \Cref{sec:new-viz}.

\subsection{Ecosystem}
\label{sec:ecosystem}

\nwk currently supports general-purpose environment managers \texttt{conda} (channel \texttt{conda-forge}, \url{https://conda-forge.org}) and \texttt{brew}
(for mac\-OS, \url{https://brew.sh})
as well as the HPC-focused manager
\texttt{spack} (\url{https://spack.io}).
While brew and conda packages are provided as pre-built binary packages per default, \texttt{spack} is useful for creating optimized binaries for a complete toolchain.
\texttt{conda} is one of the more popular package managers for data scientists, having the majority of analysis tools available while providing a low entry barrier for setting up working environments.
Prominent frontends for creating analysis workflows like
variants of Jupyter (such as the notebook server or the
Jupyter hub platform~\cite{DBLP:conf/xsede/StubbsLPCZFVFD20,DBLP:journals/cse/Barba21,DBLP:journals/cse/FangohrKD21,DBLP:conf/msr/PimentelMBF19})
use \texttt{conda} as their main channel of distribution.

Besides supporting package managers, \nwk is available on all major platforms (Linux, macOS, and Windows OS) as a pre-built wheel-package for x86\_64 architectures, Linux aarch64 and Apple arm64 via PyPi. Hard dependencies for \nwk are kept minimal in order provide easy integration into established toolchains. The binary version only needs \texttt{scipy} and \texttt{numpy}\cite{DBLP:journals/nature/HarrisMWGVCWTBS20,DBLP:journals/corr/abs-1907-10121}. Every other dependency is either bundled (like \texttt{tlx}) or optional (like \texttt{pandas} and \texttt{tabulate}).

More complex data analysis workflows, often spanning different fields of expertise, introduce additional requirements and complexity. Furthermore, from the maintainer's perspective, the ever-growing list of ecosystems makes it harder to keep up. Often, a certain required package or version is only available for a certain platform or manager tool, whereas composing an environment with different tools is not always possible.
Hence, \nwk is also available as a Docker image providing a complete environment with all optional and non-optional dependencies and software. The image is based on a Jupyter base image, which itself is based on an LTS version of Ubuntu Linux. It includes documentation and interactive tutorials, accessible via a pre-configured Jupyter-Lab instance. 


\section{\nwk in the Cloud}
\label{sec:cloud}

There are some use cases like multi-user setups or on-demand services, which can be a complex endeavor itself when starting with container images as the only tool. We deal with this by using the Dockerfile of \nwk as a basis for providing an on-demand cloud service based on a Kubernetes~\cite{DBLP:journals/cloudcomp/Bernstein14b,burns2019kubernetes} infrastructure.
By having \nwk available as a cloud-based service, domain experts can try out the complex environments with a low barrier and in a collaborative manner.
An application hosted on the cloud server enables the domain expert to interactively gain an intuitive understanding of the network analysis tools
that \nwk offers and how they map onto the domain problem under
investigation.
In contrast to public services such as Binder (\href{https://mybinder.org}{https://mybinder.org}), hosting a pre-bundled software environment leaves more flexibility when dealing with
features such as multi-users or mounting external data sets. 
For many use cases, publicly available services are also not
sufficient regarding CPU and memory requirements.
This is especially true when -- like in the
use case presented in this paper
-- interactivity or extensive software setups spanning multiple
scientific areas are part of the requirements.
In the following, we describe our approach of an on-premise state-of-the-art Kubernetes infrastructure and how it can be utilized for data science problems.

\subsection{Infrastructure}

\begin{figure}[tb]
\centering
\includegraphics[width=.45\textwidth]{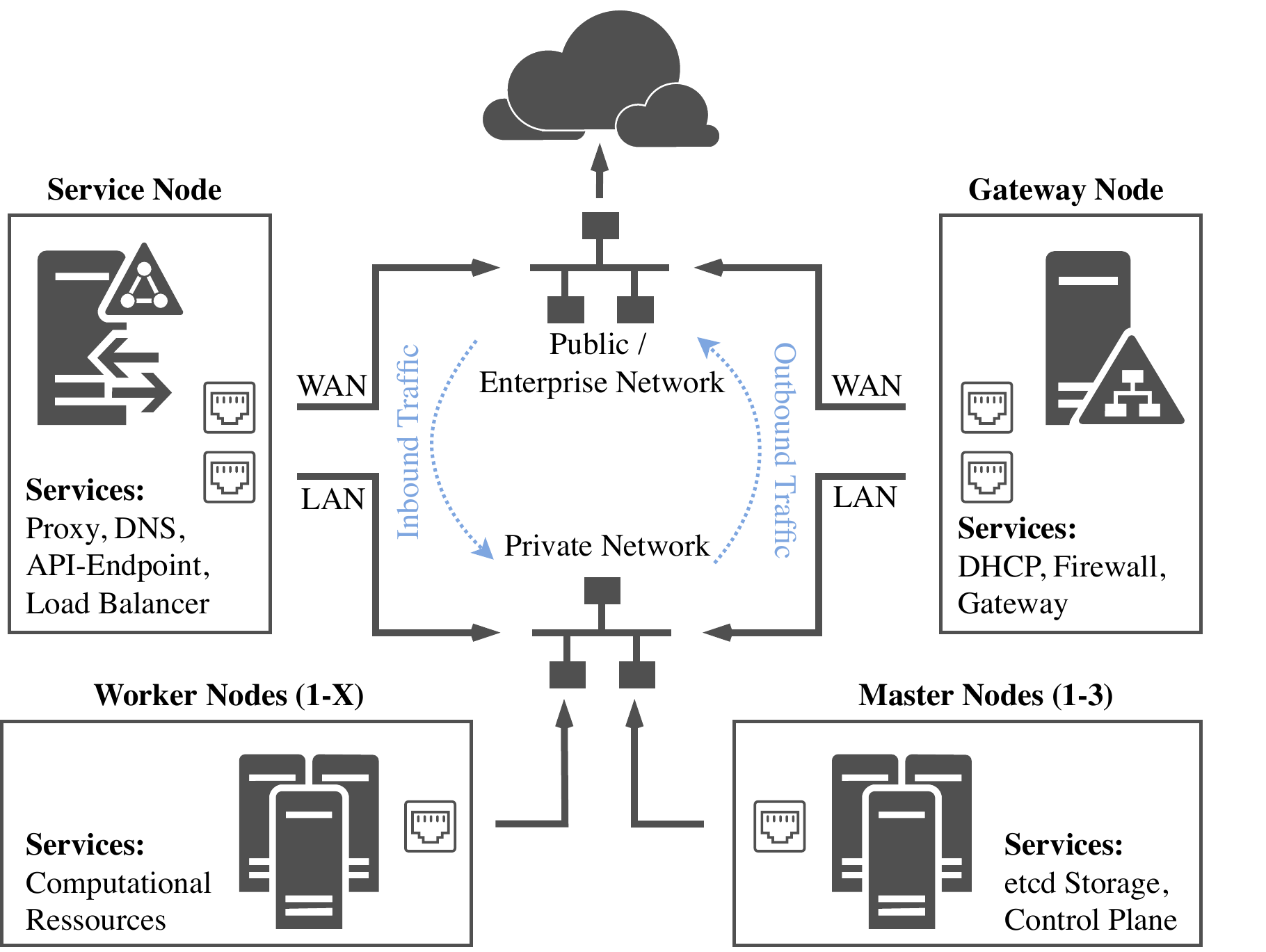}
\caption{Setup for a high-availability Kubernetes cluster in a private virtual network. Availability of control plane and storage is provided by three identical master nodes, whereas external access is managed by a load balancer on a service node, being part of both the private and a public routable network. A gateway node handles outgoing packages.}
\label{fig:kubernetes-network}
\end{figure}

The cloud-based implementation of \nwk is currently relying on a RedHat OpenShift 4.9 high-availability cluster~\cite{DBLP:conf/cascon/LinzelZFLD19},
embedded on-premise in a university network.
For the general infrastructure setup, we created the cluster based on established rules\footnote{\href{https://kubernetes.io/docs/setup/production-environment/}{https://kubernetes.io/docs/setup/production-environment/}} for production environments embedded in a private and isolated network.
The three master and two worker nodes span a high-availability Kubernetes cluster, providing several instances of etcd key-value stores, control-plane instances and computing resources. 
An additional service node acts as a reverse proxy and load balancer for the public services, as a DNS for the private network spanned by all participants and an entry point for the Kubernetes API. 
All traffic into the cluster is routed through this node. The gateway node handles the reverse route from within the cluster to WAN, equipped with an additional ACL-based firewall and filter mechanism to monitor traffic. This approach of embedding a HA-cluster helps to secure the Kubernetes nodes, providing good scalability in terms of adding more resources.

Hardware requirements for all except the worker nodes can be kept at a moderate level. Master and service nodes need at least 4~CPUs and 16~GB of memory in order to provide good scalability, which are also the specifications used in our current implementation.
Worker nodes should always scale with the desired use case, which for \nwk means to deal with data analysis workflows in one or several container instances. 
The network stack involves at least a Jupyter instance for the service endpoint and data is likely duplicated during the workflow process. Hence, memory to manage data structures and web frontends is the most important requirement, followed by CPU cores for parallel processing. 
The benchmarks presented in this paper were conducted with a limit of 10 vCores and 16~GB of memory for each instance, which has proven to work well for our use case.

\begin{figure}[tb]
\centering
\includegraphics[width=.45\textwidth]{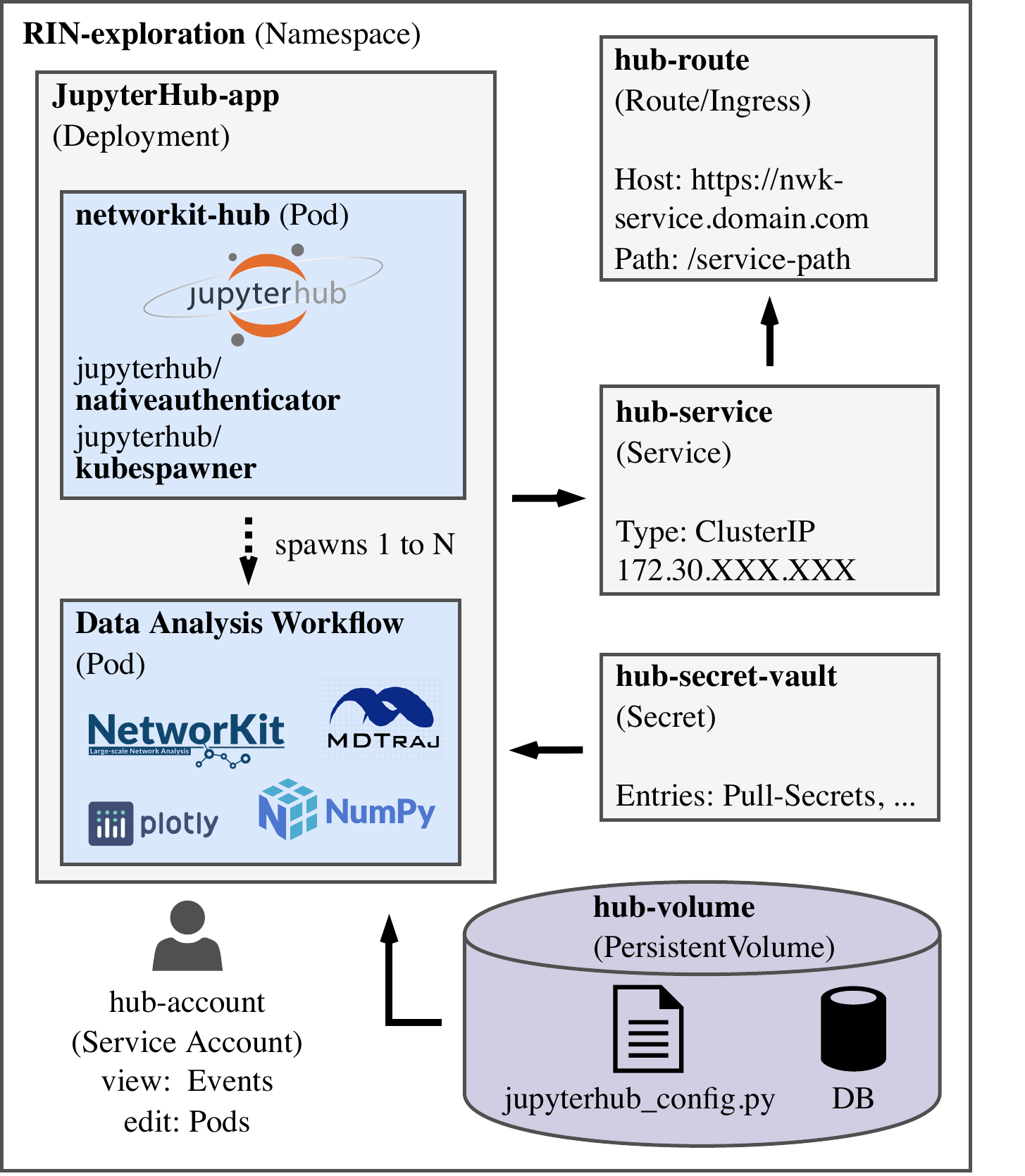}
\caption{Kubernetes entities used to create a service for a typical data analysis workflow based on Jupyter notebooks. A namespace contains all relevant entities, including a central deployment and all necessary additions (secrets, volume, service account). Exposing the deployment is done via a service + ingress/route definition.}
\label{fig:kubernetes-service}
\end{figure}

\subsection{Service Definition}

Setup of the public service for the end user is done by defining Kubernetes entities as shown in \Cref{fig:kubernetes-service}. The process is following state-of-the-art rules by creating a cluster internal deployment resource and enabling external access via a service and ingress/route resource. 
For the deployment resource, a customized JupyterHub Docker container is built, adding necessary authentication and instance spawner plugins.
While the authentication plugin can be chosen arbitrarily based on the hosting infrastructure and use case, it is recommended to use \texttt{KubeSpawner} for the spawner plugin.
This allows JupyterHub to start on-demand additional Kubernetes pods hosting the user instances, with proxy access managed within the original deployment.
Realizing pod spawning from within a running pod in a Kubernetes cluster is done via connecting a service account (SA) to the JupyterHub deployment. This SA has to be granted at least view permissions for Kubernetes events and permissions to spawn, list, and delete pod resources. 
The JupyterHub instance is created in its own namespace, so that different service endpoints cannot interfere with each other.
Therefore, also the permissions for the SA can be local to that namespace in order to minimize to risks in case of a security-related incident. 
Persistence concerning configuration and accounting is achieved by adding physical volumes (PV) and physical volume claims (PVC), containing a pre-configured \texttt{jupyterhub\_config.py} file and user database. The config contains information such as image name and pull secrets for user instances or cgroup limits. 

Reaching the service in the private cluster is done by attaching a domain/IP to the public interface of the service node. 
The reverse proxy connects the public WAN interface with the cluster network and forwards a service query on http/https port to one of the worker nodes based on a source balanced policy.
Each node runs a replicated cluster-internal second reverse proxy, which has a prefix-based routing.
Based on the URL defined ingress/route entity, the reverse proxy forwards the package to the pod on the appropriate worker node. 

In order to create a certain use case, there are two possibilities. (i) The pod can be exchanged or (ii) another namespace with its own JupyterHub instance can be created. Both strategies can be achieved easily, making this general approach versatile and supported by all variants of cloud providers. A generalized version of the service definition used in this paper is available on Github\footnote{\href{https://github.com/networkit/networkit-cloud}{https://github.com/networkit/networkit-cloud}}.

As mentioned in \Cref{sec:ecosystem}, the underlying workflow analysis pod for this use case is based on a \nwk container. For the use case of analyzing RINs, a custom container is built; it provides a ready-to-use environment with packages for network and data workloads, protein processing, and visualization, tailored to be used in a Jupyterlab frontend. 
Entry point for notebooks is an automatically activated conda environment with over 200 packages, handling both setup and Python APIs\footnote{A link to a playground for review purposes can be found at the end of the paper.}.


\section{Residue Interaction Networks (RINs) for molecular dynamics simulations of proteins}
\label{sec:rin}

Here, we briefly introduce the application domain of protein MD simulations. We describe how network analysis can be applied to better understand the structural dynamics of proteins from large MD data sets. 
Proteins are complex and dynamic systems made up of a linear sequence of amino acids. 
The interactions between their residues -- the side chains of the amino acids -- determine the three-dimensional structure of the proteins \cite{Shcherbinin2019AnalysisNetworks}. 
How this structure changes over time determines the function of the protein. These structural dynamics of proteins play a central role in biological processes like protein folding \cite{Lindorff-Larsen2011}, cellular signaling \cite{Berg2020}, or protein-protein interactions \cite{Sikora2021ComputationalProtein}.

Molecular dynamics simulations offer an increasingly powerful computational tool to investigate these processes at temporal and spatial resolutions that are hard or impossible to attain experimentally \cite{Schlick2021BiomolecularTechnology}. 
In an MD simulation, a protein is simulated as it evolves under the forces of classical mechanics. The data sets resulting from these simulations are large, high-dimensional time series of protein structures called trajectories. The individual frames of the trajectory contain the Cartesian coordinates of all atoms in the system \cite{Hollingsworth2018Molecular}. 
An MD simulation of a protein can have millions of frames and thousands of dimensions. The handling and analysis of MD data is commonly performed in analysis pipelines built in Python, chiefly due to the large community working on and with Python tools for data transformation, machine learning applications, and visualizations \cite{Noe2020MachineSimulation, Ceriotti2019UnsupervisedUnderstanding, Glielmo2021UnsupervisedData}.

However, extracting insights on the protein function from these time series of atom coordinates is challenging. A powerful approach to make the analysis of protein structural dynamics more tractable is to describe the three-dimensional structure of the protein as a graph.
Graphs are a natural formalism to capture protein structure \cite{DiPaola2013} and they make it possible to treat the protein system with a rigid, mathematical formalism and apply an established set of algorithms to better understand it.
To analyze entire trajectories from MD simulations, each frame of the simulation can be translated into a graph representing the underlying protein structure. 
Such a graph is referred to as an amino acid network, side-chain network, or residue interaction network (RIN) \cite{Eds}. In a RIN, the nodes of the graph represent the amino acids and the edges represent their interactions. 

\begin{figure}[tb]
\centering
\includegraphics[width=\columnwidth]{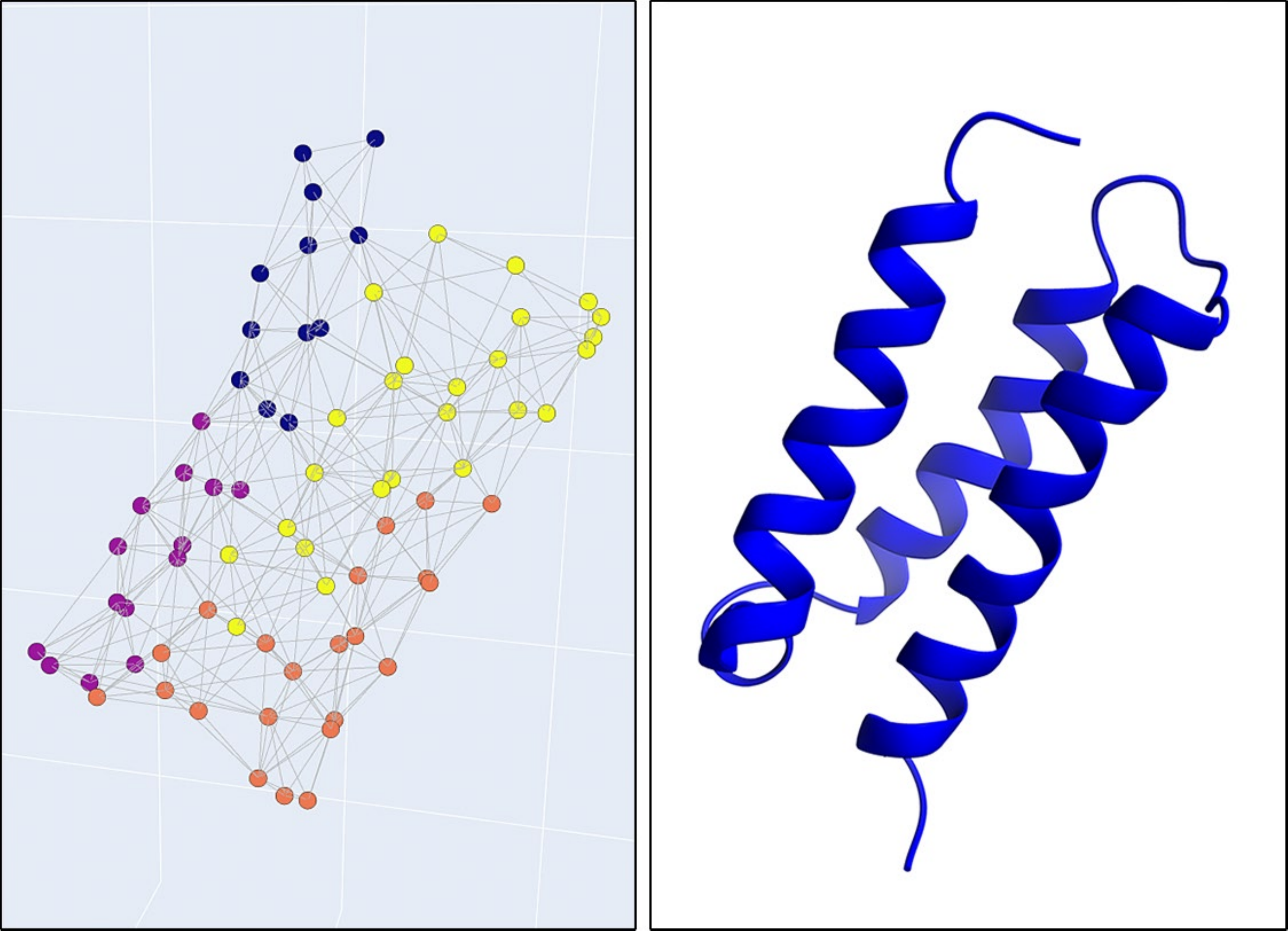}
\caption{Left: 3D-plot of the RIN of $\alpha$3D at a minimum distance cut-off of  \SI{4.5}{\angstrom}, colored by communities found by PLM community detection. Right: Protein structure of $\alpha$3D. The secondary structure elements ($\alpha$-helices) are reflected in the community structure of the RIN.}
\label{fig:RIN-structure}
\end{figure}

There is a host of methods to translate proteins into RINs. A common way is based on the spatial distances between the residues in the protein structure. Depending on the question, the residue-residue distance can be determined in different ways, such as the distance between the C-$\alpha$ atoms of each residue, the centers of mass of the residues, or the distance between whichever two atoms are closest to each other in the respective residues (minimum distance). Commonly, the residues (nodes) are considered to be interacting (connected by an edge), if a residue-residue distance is within a given cut-off distance. The cut-off usually varies between \SIrange{4}{8.5}{\angstrom}, depending on the way the distance was calculated and the scientific question \cite{yan2014construction, k2015modeling}. The resulting RINs are commonly unweighted and undirected graphs representing the protein structure. An example for a RIN constructed from the fast folding protein $\alpha$3D is shown in \Cref{fig:RIN-structure} next to its three-dimensional structure. 

Once the protein structure is suitably translated into a graph, many different network-analytic methods can be applied to better understand the protein structure and how it changes over time.
The network analysis of RINs has been applied to answer a plethora of questions on the relationship between protein structure and function, e.g. analyzing the role of mutations in the SARS-CoV-2 spike protein~\cite{Verkhivker2021DynamicSwitches}.
Various node centralities have been used to identify functionally important residues in protein structures, such as protein-protein interfaces and active sites. 
Often several residue centralities are combined with each other or with additional information on the residues, such as surface accessibility, to improve the results~\cite{Jiao2017PredictionNetwork, DBLP:journals/bmcbi/CheaL07, Amitai2004NetworkResidues}. 
Community detection algorithms have been applied to find structural modules in the protein \cite{Grant2019ModularDetection, Tasdighian2014ModulesSolutions}, to understand regulatory mechanisms \cite{DelSol2007ModularLinkages} and to investigate information flow within the protein \cite{Stetz2017ComputationalCommunication}. 

When starting the analysis of a new protein system, it can be difficult to decide which translation algorithm and which network analysis algorithm is best suited to answer the underlying biological question about it \cite{Halder2020SurveyingCase, Eds}. There is not one unique recognized way to best use network analysis for investigating a protein. And since each protein system is different and has different underlying scientific questions, it is vital to gain an understanding of the system and how network-analytic tools capture it.
For example, it has been shown that the choice of the distance criterion can influence which secondary structure features are emphasized \cite{DaSilveira2009ProteinProteins} and changes in the distance cut-off can drastically alter the RIN topology, e.g. influencing the number of hubs and connected components \cite{Viloria2017AnMass}. Different centralities can indicate different roles for individual residues. A high betweenness centrality for a residue could indicate that the residue plays an important role in a protein-protein interface \cite{Jiao2017PredictionNetwork}, in information flow through the protein \cite{Stetz2017ComputationalCommunication}, or in maintaining a specific fold \cite{Haratipour2019NetworkTopology}. 
A high closeness centrality, in turn, could indicate a role in the protein's active site or a ligand-binding site \cite{Amitai2004NetworkResidues, delSol2006ResidueFamilies, DBLP:journals/bmcbi/CheaL07}. 
There are many applications of community detection algorithms on RINs to investigate various questions, and different community detection algorithms can differ in the structural features they capture~\cite{DiPaola2015}. 

\section{Visualizing RINs with \nwk \\ and Jupyter Notebooks}
\label{sec:new-viz}

To get a good understanding of how the network-analytic measures map onto the protein system of interest and how they change with different cut-off values for translating the protein structure into an RIN, it is crucial to be able to visually explore the parameter space for the cut-off and the network measures.
Since proteins are dynamic structures that change during the course of a simulation, it is also crucial to be able to perform these analyses on all protein structures of the simulation trajectory. 
This can convey an understanding how the RIN topology and corresponding network measures change over time and with different cut-off values. 

An application that makes it possible to interactively vary all relevant parameters and visually follow changes in RIN topology in real time can give an intuitive understanding of the relationship between the protein system and the network description. In addition not only the topology, but also the geometric structure resembling the protein folding is of interest, making a 3D representation more suitable than 2D. To efficiently explore the parameter space outlined above, a tool that can support such a visualization task has several requirements.

\begin{enumerate*}[label=\lbrack R\arabic*\rbrack]
  \item It should include all network analysis tools of interest, such as centralities and community detection algorithms.
  \item It should seamlessly integrate into a Python analysis workflow, such that it can be integrated easily with any upstream data transformation tasks (e.g. data filtering and selection) or further downstream analysis (e.g. applying machine learning algorithms on the extracted network features).
  \item  It should calculate and visualize all network features fast enough to facilitate at least near real-time, interactive exploration of the parameter space.
  \item It should accommodate the simultaneous manipulation of the parameters of interest (cut-off for graph translation, network metric and trajectory frame) and the visualization of the resulting network features and topologies in one user interface.
  \item It should be able to handle large, high-dimensional data sets from MD trajectories.
\end{enumerate*}

\subsection{Application Design}

\nwk does not directly
implement any visualization GUI (although it includes some
graph drawing algorithms). Instead, we rely on well-known and established
tools to draw the GUI of our visualization widgets.
We provide adapters that allow these tools to take \nwk graphs as
input.
With the release of version 3.2, a streaming client for \gephi\cite{DBLP:conf/icwsm/BastianHJ09} has been implemented. 
\gephi is a well-known and powerful tool for the visualization of networks providing an interface for static and dynamic graphs and algorithms for graph drawing such as \texttt{Force Atlas 2}\cite{DBLP:conf/icwsm/BastianHJ09} or \texttt{Fruchterman Reingold}\cite{DBLP:journals/spe/FruchtermanR91}. 
However, it is written in Java using Swing for its GUI. Even though techniques for converting classical Swing frontends to web technologies (e.g. WebSwing)\footnote{\href{https://www.webswing.org}{https://www.webswing.org}} have been developed over the past years, there is no straightforward solution to directly embed Java programs into Jupyter Notebooks and therefore Python frontends.
Another approach is to use common plotting tools based on HTML5, SVG and Javascript and create adapters to integrate them into Jupyter notebooks. \nwk implements two modules \texttt{csbridge} (2D graphs) and \texttt{plotlybridge} (2D and 3D graphs), which are based on \texttt{ipywidget} and act as such an adapter for embedded graph visualization. These widgets use external Python packages \texttt{ipycytoscape} and \texttt{plotly}, which has previously been used to visualize RINs \cite{Saini2021}. For the scope of this paper and the requirements for RIN-exploration, we will focus on the module \texttt{plotlybridge}, while in principle \texttt{csbridge} follows a similar approach.

Like \gephi, \texttt{Plotly} (\href{https://plotly.com}{https://plotly.com}) is a well known tool for visualization; it is, however, focused on plotting charts and mappings in both 2D and 3D. As a consequence there are no graph drawing algorithms available -- the user has to provide coordinates for available data points. For the Python interface, each chart in \texttt{Plotly} is represented by a \texttt{plotly.graph\_objects.FigureWidget}, which is a custom \texttt{ipywidget} usable for embedding in more complex GUIs. One or more data sets can be added to the widget by calling \texttt{add\_traces()}.

\begin{lstlisting}[language=Python, caption={Initialization of plotlybridge-module (shortened).}, label={lst:plotlybridge-init}]
import networkit as nk
import plotly.graph_objs as go

# plotlyWidget-functionality (shortened)
def plotlyWidget(G, scores):
    maxLayout = nk.viz.MaxentStress(G, 3, 3)
    maxLayout.run()
    coordinates = maxLayout.getCoordinates()
    
    figWidget = go.FigureWidget()
    nodeScatter, edgeScatter = go.Scatter3d(...), go.Scatter3d(...)|\label{line:init}|
    figWidget.add_traces(nodeScatter, edgeScatter)
    figWidget.layout = go.Layout(...)
    ...
    return figWidget

# Graph init + score-computation
G = nk.readGraph("karate.graph", nk.Format.METIS)
betCen = nk.centrality.Betweenness(G)
betCen.run()
scores = betCen.scores()
plotlyWidget(G, scores)
\end{lstlisting}

\begin{figure}[tb]
\centering
\includegraphics[width=.50\textwidth]{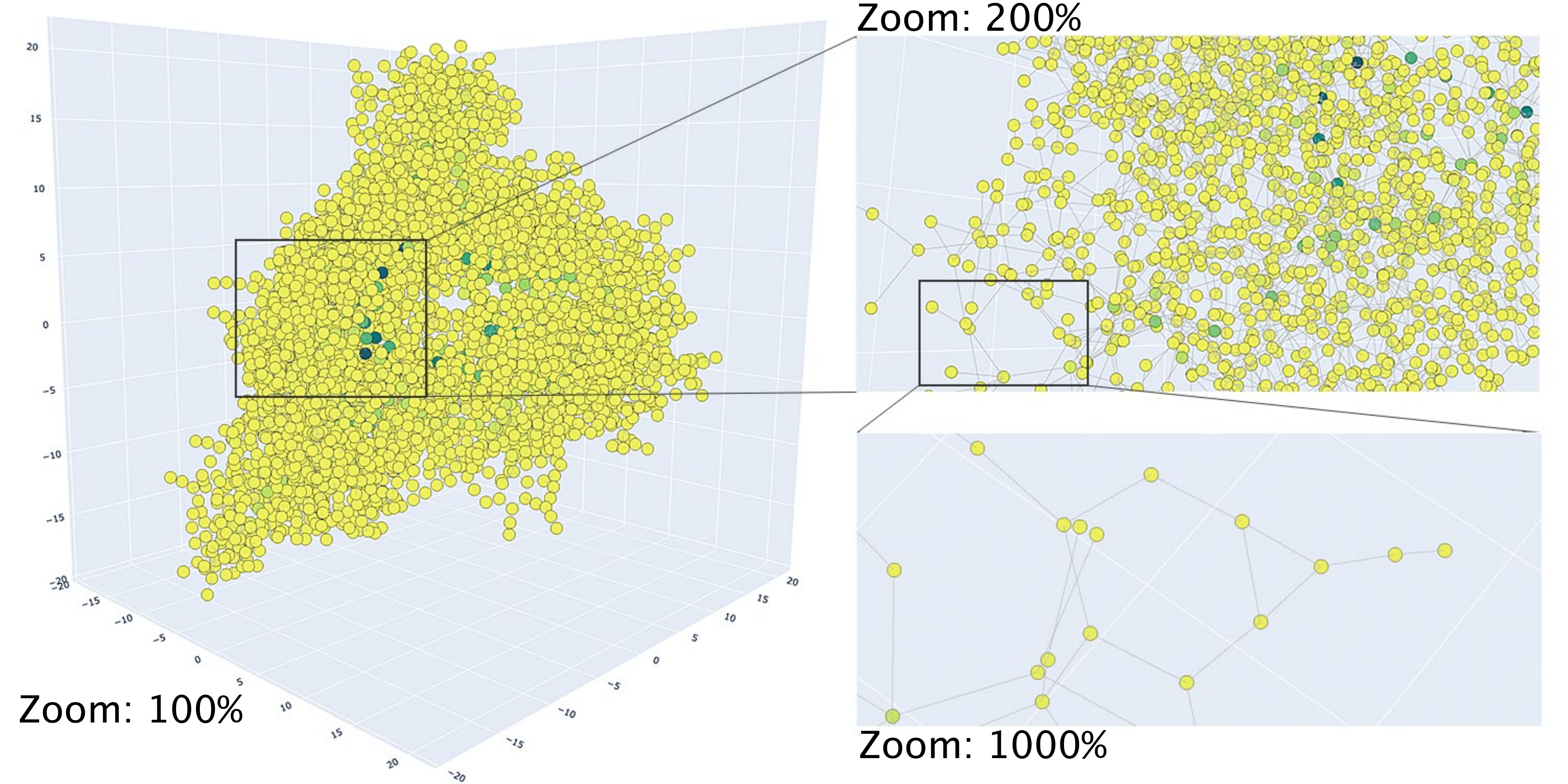}
\caption{3D-plot generated by \texttt{plotlybridge}-module, showing a graph with 4941 nodes and 6594 edges. With a few lines of codes, the nodes display overlay information like centrality measures by using color-palettes.}
\label{fig:plotlybridge-singleplot}
\end{figure}

\begin{figure*}[tb]
\centering
\includegraphics[width=\textwidth]{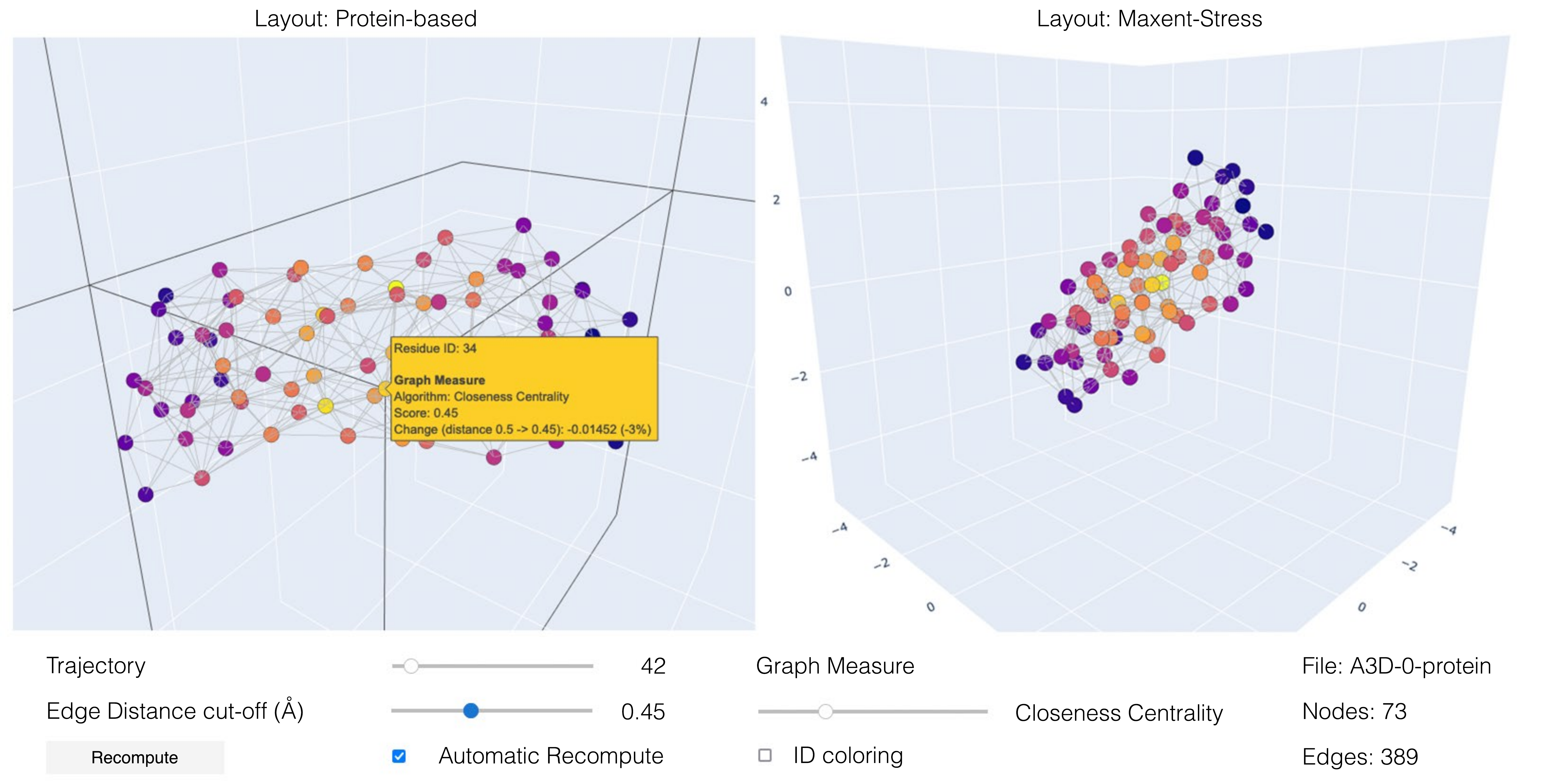}
\caption{GUI for analyzing RIN using \texttt{plotlybridge} consisting of several \texttt{ipywidgets}. Top: Dual interactive 3D-graph with Protein-based layout (left) and Maxent-Stress layout (right) for a snapshot/trajectory during simulation. Coloring of the nodes is done with a spectral color palette (blue - red), whereas each color is defined by Closeness-value of the node in the network. Bottom: Slider for selecting a trajectory frame, edge cut-off distance and network measure.}
\label{fig:plotlybridge-dualplot}
\end{figure*}

\nwk's \texttt{plotlybridge} is implemented as a Python function, called by handing over a graph object \texttt{G} and an algorithm with the ability to compute node scores. Before the visualization figure is created, 3D coordinates have to be generated. For this, the Maxent-Stress algorithm from \nwk~\cite{DBLP:conf/esa/WegnerTSM17} is used. The main intention of this algorithm is to solve an optimization problem that computes the three-dimensional structure of biomolecules, which makes it also suitable for computing the layout for our current use case.  \Cref{lst:plotlybridge-init} shows an abstraction of how the function works. In Line~\ref{line:init}, two \texttt{Scatter3D}-objects are created, one containing the nodes represented by 2D-circle shapes and the other containing the edges represented by 2D-lines. Projecting 2D-shaped nodes and edges correctly, while allowing for change of perspective and zoom is done by \texttt{Plotly} itself. The \texttt{plotlybridge}-widget is finalized by defining a layout, including axis, colors and 3D camera modifications. Since generating a layout and creating the data is done in Python/C++, most of the heavy computation is done on the server side.
This allows to draw graphs with up to 50k nodes in a few seconds on commodity hardware (see \Cref{fig:plotlybridge-singleplot}).

\begin{figure*}[tb]
\centering
\includegraphics[width=\textwidth]{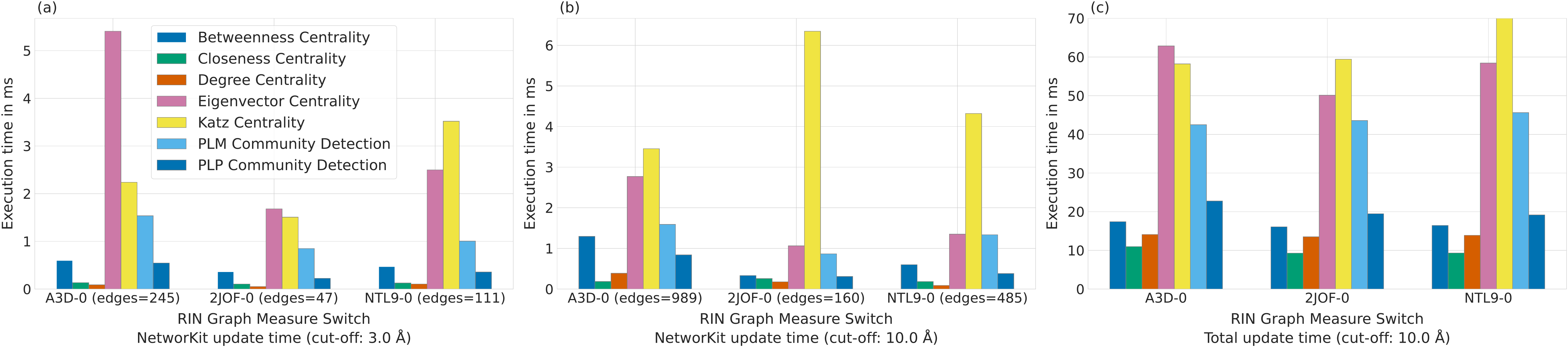}
\caption{Time (ms) it takes to recalculate popular centrality and community detection measures on different RIN-networks. $(a)$ + $(b)$ show computational time, which NetworKit takes to update the underlying network for low and high cut-off values. Higher cut-off values increase the number of edges. $(c)$ shows the time for the whole update cycle as perceived on the client. This includes updating the underlying network, the widget data-handling and the Plotly-graph.}
\label{fig:plotlybridge-data-algo}
\end{figure*}

\begin{figure*}[tb]
\centering
\includegraphics[width=\textwidth]{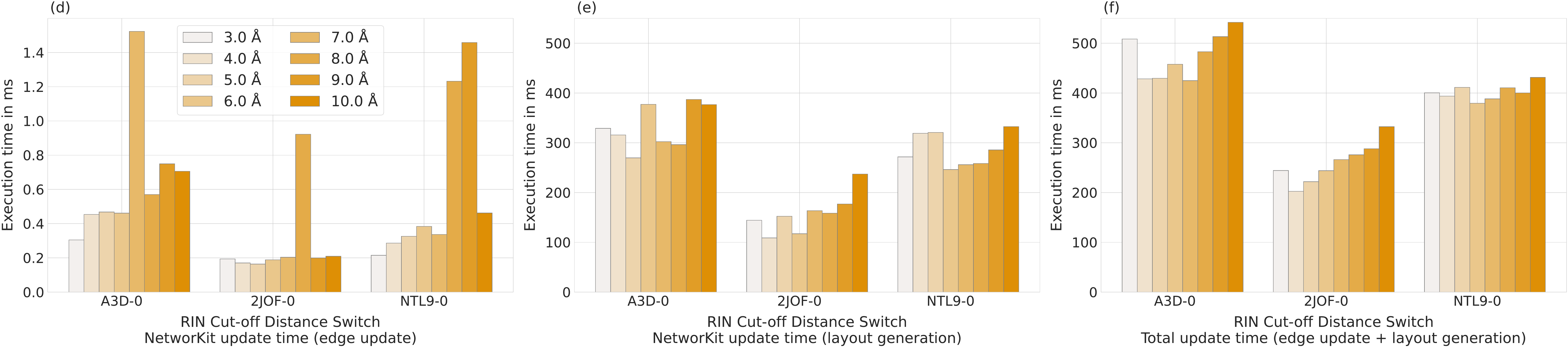}
\caption{Time (ms) it takes to switch between different cut-off distances on different RIN-networks. Each switch consists of an edge update and an layout generation phase. $(d)$ shows the computation time \nwk takes to update edges in the RIN network. $(e)$ shows computation time \nwk takes to generate the Maxent-Stress layout. $(f)$ shows the time for the whole update cycle as perceived on the client. This includes updating the underlying network, the widget data-handling and the Plotly-graph.}
\label{fig:plotlybridge-data-cutoff}
\end{figure*}

\begin{figure*}[tb]
\centering
\includegraphics[width=\textwidth]{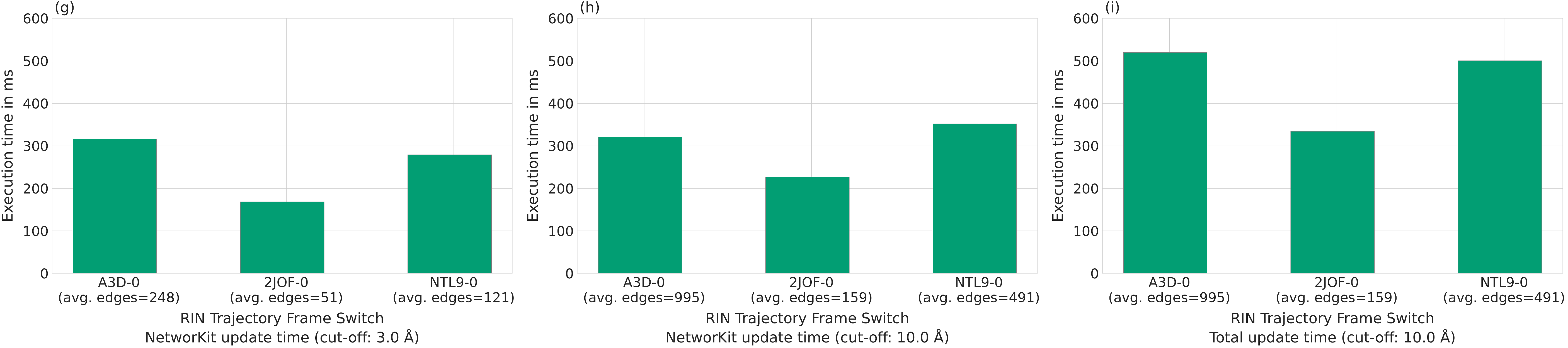}
\caption{Time (ms) it takes to switch between different trajectory frames on different RIN-networks. $(g)$ + $(h)$ show the computation time \nwk takes to update the underlying network for low and high cut-off values. Higher cut-off values increase the number of edges. $(i)$ shows the time for the whole update cycle as perceived on the client. This includes updating the underlying network, the widget data-handling and the Plotly-graph.}
\label{fig:plotlybridge-data-trajectory}
\end{figure*}

Using the Python API and the speed of \nwk, the code-snippet from the \texttt{plotlybridge} widget shows that a few lines of code can already create an interactive visual inspection tool for small and medium-sized graphs. Adding information with respect to different network metrics like centrality measures or community structures is easily achieved by node coloring and text-box displays. Since the widget is implemented as an \texttt{ipywidget}, we achieve this by stacking the basic version of the \texttt{plotlybridge}-widget into a more complex GUI and adding functionality for transforming node properties to color scales. In addition to the widget of the use case in \Cref{sec:rin}, we add more functionality tailored for inspection of proteins represented by RINs. 

\subsection{Benchmarking Results}

The final software version\footnote{A link to a playground for review purposes can be found at the end of the paper.} fulfilling the requirements as stated above is shown in \Cref{fig:plotlybridge-dualplot}. The code is also available on Github\footnote{\href{https://github.com/hu-macsy/2022-paper-RIN-visualization}{https://github.com/hu-macsy/2022-paper-RIN-visualization}}. 
We add interactive sliders, which let the user choose between different popular network measures on the fly. For smaller networks like RINs of proteins with around 100-1000 nodes, computing centrality or community properties only takes a sub-second time span in total.
Two additional sliders let the domain expert choose between different RIN trajectory frames from the simulation of a protein and different cut-off distances for translating the protein structure into an graph. 
Both values do not change the number of nodes in the network, but influence layouts and the number of edges. 
These changes in the network topology also imply changes to the node properties and therefore trigger an update of the chosen network measure. 
Additional misc.\ components in the GUI provide quality of life functions and let the domain expert choose whether re-computation is done automatically or on demand.
By storing the most recent computed node property within a buffer in the widget, it is also possible to visualize the delta between different cut-off distances or trajectory frames.

On the upper half of the widget, two side-by-side 3D graph-plots show the same RIN, but with different layouts. The left plot shows the real conformation of the protein with the node positions determined by the positions of the C-$\alpha$-atoms in the protein. The right plot shows the Maxent-Stress layout, computed by \nwk. Having the two plots side-by-side adds an additional layer for decision making, in particular on 1) how the chosen network metric matches the protein structure (e.g. whether putative or known functionally important residues are identified) and 2) how the cut-off affects the topology of the RIN (e.g. if it is far off from the actual protein structure, the cut-off may be unsuited). The Maxent-Stress layout can also help explain or highlight some network features that may not be immediately obvious from the protein-structure-based layout, such as community structure across or within secondary structure elements.

Executing the widget is done inside the custom built container on the cloud infrastructure (see \cref{sec:cloud}) using the lab interface from Jupyter. While all network data manipulation is handled on the cloud infrastructure backend by \nwk, the protein to RIN conversion is done by MDtraj\cite{mcgibbon2015mdtraj}; supporting data structure and logic in Python use \texttt{numpy}. For different simulations of protein folding, which are commonly used by domain experts to assess algorithms for the analysis of protein MD data \cite{Lindorff-Larsen2011}, we can see in \Cref{fig:plotlybridge-data-algo,fig:plotlybridge-data-cutoff,fig:plotlybridge-data-trajectory} that the whole computation pipeline takes sub-seconds, sometimes even single-digit milliseconds time. 
Some of the results include updating the output of Jupyter notebooks, therefore client specifications also play a role here. 
All benchmarks were carried out with Firefox 96.0 on a MacBook Pro M1 with 16GB of main memory. 
From \Cref{fig:plotlybridge-data-algo} $(a)$ and $(b)$ we can see that calculating the network measures takes only single-digit milliseconds, whereas a complete update of the widget (subplot $(c)$) takes 10x more time. 
Depending on the network measure, the result is suitable for fluent animation or video playback (\SIrange{24}{60}{fps}). 
Most of the time is spent in updating the visuals, which is natural given the necessary update of the underlying DOM-elements. For switching between different cut-off distances (see \Cref{fig:plotlybridge-data-cutoff}) and trajectory frames (see \Cref{fig:plotlybridge-data-trajectory}), the situation is somewhat reversed. 
Most of the overall update time is spent during updating the underlying network in \nwk, whereas the functionality for switching cut-off distances and trajectory frames is similar, leading to nearly identical running times for equal cut-off distances. 
Both routines consist of adding/removing edges and recomputing the Maxent-Stress layout phase. From \Cref{fig:plotlybridge-data-cutoff} $(d)$ and $(e)$ we can see that recomputing the layout takes the majority of the time with around \SIrange{300}{400}{\milli\second}.
The complete update adds roughly \SI{100}{\milli\second} for a cut-off change event (see \Cref{fig:plotlybridge-data-cutoff} $(f)$) and \SI{200}{\milli\second} for a frame change event (see \Cref{fig:plotlybridge-data-trajectory} $(h)$). This is expected, since the number and position of nodes for the protein layout remain equal between different cut-off distances. Therefore for this event, the protein-based layout graph only needs to update the DOM-elements representing the edges. 
When a trajectory frame is changed, though, the protein structure and therefore the position of nodes in the RIN have changed -- making a complete update to all visualization-related DOM-elements necessary. 
Note that the maximum amount of time it takes to update the widget is occurring on changing the trajectory, while having selected a network measure.
The related update functions are called subsequently, leading to a total loop time of up to approx. \SI{600}{\milli\second} for networks with around 1000 edges (see \Cref{fig:plotlybridge-data-trajectory} $(i)$).
As long as the resource provisioning does not create bottlenecks on the cloud infrastructure, the server-based performance metrics are stable and provide real-time results.


\section{Related Work}
\label{sec:related-work}

\paragraph*{Other graph toolkits}
Several network analysis tools
are publicly available and they offer a wide
range of trade-offs between performance, usability, and number of features.
NetworkX~\cite{hagberg2008exploring} is a popular Python package, can be installed
and used with very little effort, and offers a wide range of algorithms and features.
It is, however, written in pure Python, and the performance is thus limited.

Similarly to \nwk, packages such as graph-tool~\cite{peixoto_graph-tool_2014},
igraph~\cite{csardi2006igraph}, or SNAP~\cite{DBLP:journals/tist/LeskovecS16}
implement performance-aware algorithms in C++ (in the igraph case: C) and
provide bindings for Python. This architecture offers good usability without
sacrificing performance -- see \Cref{sec:ecosystem}. However, it requires to
compile the C/C++ core, which is an additional burden to be considered.

Other tools such as Ligra~\cite{DBLP:conf/ppopp/ShunB13},
Stinger~\cite{DBLP:conf/hpec/EdigerMRB12},
Hornet~\cite{DBLP:conf/hpec/BusatoGBB18}, and others (see
Ref.~\cite{DBLP:journals/jbd/CoimbraFV21} for an exhaustive overview) are
available only as C++ libraries, and provide a collection of high-performance
algorithms -- often much more limited compared to the aforementioned packages.

\paragraph*{Other tools for RIN analysis} 

There are numerous tools available for constructing, analyzing and visualizing RINs (see \cite{yan2014construction, Chakrabarty2016NAPSStructures, Eds} for detailed overviews). There are plug-ins for established visualization software, such as RINalyzer \cite{Doncheva2011AnalyzingStructures}, which ties in with structureViz \cite{Morris2007StructureViz:Chimera} to link Cytoscape \cite{Shannon2003Cytoscape:Networks} and the molecular structure viewer Chimera \cite{Pettersen2004UCSFAnalysis}. Standalone software like gRINN \cite{Sercinoglu2018GRINN:Simulations} and Webservers like CMWeb \cite{Kozma2012CMWeb:Methods} offer GUI-based visualization tools with a varying number of features. However, these tools offer limited capabilities when it comes to handling large data sets of full simulation trajectories with real-time responsiveness. They often require separate installation or depend on other software. To our knowledge, none of the existing tools can be integrated seamlessly into python analysis workflows, limiting the possibilities to adapt the network analysis to user-specific needs.  


\section{Conclusions and Future Work}

With a cloud based solution based on \nwk and JupyterHub,
it is possible for domain scientists to have a low barrier entry into learning how to adapt functions from network analysis to their specific problems. 
Having ready-to-use environments on cloud-hosted servers enables the user to claim resources on the fly and perform collaborative and shareable research.
Based on the nature of the technology, most of the expensive recomputation is delegated to the infrastructure, leading to responsive applications without the need for setting up toolchains or an expensive client workstation. 
Our approach of creating a cloud service is using current established and generalized techniques, which makes it easily transferable to other cloud infrastructures like public hosters. One future contribution could be a transition to automation tools such as Helm or Terraform, which would increase portability and enable resource provisioning based on a predefined use-case given by a single Dockerfile and data sets.

As shown with our visualization application for protein RINs, a fast GUI-based frontend can be achieved without considerable effort. It allows for flexible and interactive exploration of large data sets. 
A well-maintained Python API enables the adaptation to user-specific needs. With only few additions, the application can be extended by user-defined residue features, which could be used to better capture the structural features of the protein \cite{Jiao2017PredictionNetwork, DBLP:journals/bmcbi/CheaL07}. The  application could serve to explore how community detection algorithms can aid in coarse graining the protein structure, to make simulations of larger protein systems at reduced resolution computationally tractable \cite{Eds}. Finally, it is also possible to integrate the application into Machine Learning workflows, using tools from the Python data-science domain. Graph embeddings, like node2vec \cite{DBLP:conf/kdd/GroverL16} -- which is already part of \nwk\ -- or graph neural networks could be applied to reduce the complexity of the protein simulation data and better understand the role of individual residues in protein dynamics.

\vspace{0.2cm}

\textbf{Playground for review purposes:}\\
\href{https://dev.networkit.informatik.hu-berlin.de/rin-vis-review/}{https://dev.networkit.informatik.hu-berlin.de/rin-vis-review/}\\ (Username: reviewer, Password: BrotherDialectMonopoly?)

\bibliographystyle{ieeetr}
\bibliography{references, RIN_references}

\end{document}

%% file: macros.tex
\newcommand{\fmtLib}[1]{\textsc{#1}}
\newcommand{\nwk}{\fmtLib{NetworKit}\xspace}
\newcommand{\eg}{e.g.,\xspace}

\newcommand{\etal}{et al.\xspace}
\def\gephi{\fmtLib{Gephi}\xspace}

\definecolor{codegreen}{rgb}{0,0.6,0}
\definecolor{codegray}{rgb}{0.5,0.5,0.5}
\definecolor{codepurple}{rgb}{0.58,0,0.82}
\definecolor{backcolour}{rgb}{0.95,0.95,0.92}